\documentclass[12pt]{article}
\usepackage{cite}
\def\bq{\begin{eqnarray}}
\def\eq{\end{eqnarray}}

\begin{document}

\title{\bf Comment on
``Fermion production in a magnetic field in a de Sitter universe''}

\author{Nistor Nicolaevici
\\
\it Department of Physics, West University of Timi\c soara,
\\
\it V. P\^arvan 4, \it 300223, Timi\c soara, Romania
\\
Attila Farkas
\\
\it unaffiliated,
\\
\it Alte Str. 42, 89081 Ulm, Germany
}
\maketitle

\begin{abstract}

We point out that the transition probabilities used in a recent perturbative calculation of
pair creation in an external magnetic field in the expanding de Sitter space with the $in$ and
$out$ fermion states defined by the Bunch-Davies modes [C. Crucean et al., Phys. Rev. D {\bf 73}
044019 (20016)] are gauge dependent quantities. We examine the gauge variations of these amplitudes
assuming a decoupling of the interaction at infinite times, which allows to conclude that the
source of the problem lies in the nonoscillatory behavior of the fermion current in the infinite
future.

\end{abstract}

\\
\par
A fundamental principle of a quantum gauge theory is that the transition probabilities must be gauge
invariant quantities. In a recent paper \cite{cruc} a perturbative calculation was given for
fermion pair creation in an external magnetic field in the expanding de Sitter (dS) space.
The basic ingredient in the result \cite{cruc} is the curved-space generalization of the
tree-level amplitude in an external field in flat space QED, where both the initial $in$ and $out$
particle states were defined by the Bunch-Davies modes of the Dirac field. In this\, Comment we want
to point out that the amplitudes \cite{cruc} are gauge dependent quantities. As we show below, this can
be concluded by noting that in a potential of the form $A_\mu$=\,constant these amplitudes do not
vanish.\footnote{Essentially identical amplitudes were used in the calculation of pair creation in
an external electric field \cite{cruc1} and of the vacuum decay \cite{cota} in the same background.
Our observations apply with practically no modifications to the amplitudes there too.}

We will also provide an explanation for the mechanism behind  the gauge dependent amplitudes \cite{cruc}.
We will consider the gauge variations of the amplitudes assuming a decoupling of the interaction at infinite
times, and show that, in contrast to what happens in Minkowski space, in the adiabatic limit these variations
do not vanish. This will allow to conclude that the source of the problem lies in the nonoscillatory behavior of
the fermion density $\bar \psi_f \gamma^0 \psi_i$ at $t\rightarrow \infty$, which
in turn is a direct consequence of the infinite expansion of space in this limit.\footnote{This conclusion was
also presented by one of us in Ref.~\cite{nico}. The analysis there involves certain ambiguities. We
present here an exact calculation.}

\section{The transition amplitudes}

We first recall the form of the amplitudes \cite{cruc}. The line element of the expanding
dS space is
\bq
ds^2 = dt^2 - e^{2Ht} d{\bf x}^2.
\label{dsmet1}
\eq
Introducing the conformal time
\bq
\qquad \eta= -\frac{1}{H}\, e^{-Ht}, \quad
\eta \in (-\infty, 0),
\eq
the metric becomes
\bq
\quad ds^2=\frac{1}{(H\eta)^2}\,( d\eta^2 - d{\bf x}^2).
\eq
The amplitudes \cite{cruc} were derived from the general formula (all notation is conventional)
\bq
{\cal A}_{i\rightarrow f}=-ie \int d^4 x \sqrt {-g }\, \bar \psi_f\,
\gamma^{\mu} \psi_i\, A_\mu,
\quad \,\,
\gamma^\mu \equiv \gamma^{\hat \alpha} e_{\hat \alpha}^{\,\, \mu},
\label{genamp}
\eq
with the tetrad field in the Cartesian gauge $e_{\hat \alpha}^{\,\, \mu}= -(H\eta)\, \delta^{\,\mu}_{\,\alpha}$.
The initial and final wave functions were chosen to be the solutions of the Dirac equation in the
momentum-helicity basis which define the Bunch-Davies vacuum. These solutions have the following
form\footnote{We use with minor modifications the notation in Ref.~\cite{cruc}. The four-spinors
are in the Dirac representation.} ($\lambda=\pm 1/2$):
\bq
u_{\bf p,\, \lambda}(\eta, {\bf x})
&=&
\frac{\sqrt{\pi p/H}}{(2\pi)^{3/2}}
\times
(H\eta)^2
\left(
\begin{array}{c}
\frac{1}{2}\, e^{+\frac{\pi k}{2}} H^{(1)}_{\nu_-}(-p\,\eta)\,\xi_\lambda({\bf p})
\\
\lambda\,
e^{-\frac{\pi k}{2}} H^{(1)}_{\nu_+}(-p\,\eta)\,\xi_\lambda({\bf p})
\end{array} \right)
e^{i{\bf p} {\bf x}}, \label{umod}
\eq
and
\bq
v_{\bf p,\, \lambda}(\eta, {\bf x})
&=&
\frac{\sqrt{\pi p/H}}{(2\pi)^{3/2}}
\times
(H\eta)^2
\left(
\begin{array}{c}
-\lambda\, e^{-\frac{\pi k}{2}} H^{(2)}_{\nu_-}(-p\,\eta)\,\eta_\lambda({\bf p})
\\
\frac{1}{2}\,e^{+\frac{\pi k}{2}} H^{(2)}_{\nu_+}(-p\,\eta)\,\eta_\lambda({\bf p})
\end{array} \right)
e^{-i{\bf p} {\bf x}}, \label{vmod}
\eq
where $\xi_{\lambda}(\bf p)$ and $\eta_\lambda(\bf p)$ are the usual helicity two-spinors and
\bq
k = \frac{m}{H}, \quad
\nu_\pm= \frac{1}{2} \pm i\frac{m}{H},
\eq
with $m$\, the mass of the fermion field. The solutions (\ref{umod}) and (\ref{vmod}) in the infinite past
behave as $u_{\bf p,\, \lambda} \sim e^{-ip \eta}$, $v_{\bf p,\, \lambda}\sim e^{ip \eta}$, so that
they can be identified as positive and negative energy modes. In the calculation \cite{cruc} the same set
of modes was assumed (atypically) to describe the particle states both at the infinite past and infinite
future. With this convention, the amplitudes for pair creation from the initial vacuum are
\bq
{\cal A}({\bf p}, {\bf p}^\prime)_{\lambda \,\lambda^\prime}
=-ie \int d^4 x \sqrt {-g }\, \bar
u_{\bf p,\, \lambda}\,
\gamma^\mu v_{\bf p^\prime,\, \lambda^\prime} A_\mu.
\label{a2}
\eq
The potential $A_\mu$ was taken to be the conformal transform of the standard vector potential
associated to a static dipole magnetic in flat space, which is of the following form:
\bq
A_0=0, \quad A_i=A_i({\bf x}).
\eq
For our discussion below it will be sufficient to focus on such potentials.

Let us write explicitly the amplitudes (\ref{a2}). Notice that in all expressions above the dependence
on $\eta$ can be factored out, so that the integrals with respect to ${\bf x}$ can be separately performed.
It is convenient to organize the result in the following form ($\sigma^i$ are the Pauli matrices):
\bq
{\cal A}({\bf p}, {\bf p}^\prime)_{\lambda \,\lambda^\prime}=-ie
\hat A_i
({\bf p}+{\bf p}^\prime) [\,\xi^+_\lambda({\bf p})
\sigma^i\,
\eta_{\lambda^\prime}({\bf p}^\prime)] \, F (p, p^\prime)_{\lambda \lambda^\prime},
\label{AFgen}
\eq
where we introduced
\bq
\hat A_i({\bf q})=\frac{1}{(2\pi)^3}\int d^3{\bf x}\,
A_i({\bf x})\, e^{-i{\bf q} {\bf x}},
\label{foupot}
\eq
and ($\sigma \equiv 2\lambda= \pm1$)
\bq
F(p, p^\prime)_{\lambda \,\lambda^\prime}=f_+(p, p^\prime)-\sigma \sigma^\prime f_-(p, p^\prime),
\label{defF}
\eq
\bq
f_\pm(p, p^\prime)=\frac{\pi}{4} \sqrt{p p^\prime}\,
e^{\pm k\pi}\int_{-\infty}^0 d\eta\, \eta\,
H_{\nu_\pm}^{(2)}(-p\,\eta)  H_{\nu_\pm}^{(2)}(-p^\prime\eta).
\label{fds}
\eq
The integrand above behaves as $\sim e^{i(p+p^\prime) \eta}$ at $\eta\rightarrow -\infty$, but this can be
remedied in the usual way by introducing a convergence factor $e^{-\epsilon\vert \eta\vert}$ with
$\epsilon \rightarrow 0$.

\section{Gauge dependence of the amplitudes}

Let us consider in Eq. (\ref{AFgen}) a constant potential
\bq
A_i({\bf x})=\mbox{constant}.
\label{purgau}
\eq
In these conditions the field strength $F_{\mu\nu}=\partial_\mu A_\nu-\partial_\nu A_\mu$ identically
vanishes, so that the potential is a pure gauge. Gauge invariance requires then the amplitudes to vanish.
Let us look more closely at Eq. (\ref{AFgen}) in this case. The Fourier transform of Eq. (\ref{purgau})
is the delta function
\bq
\hat A_i({\bf  q}) =\delta^3({\bf  q})\,A_i,
\eq
so that the amplitudes are
\bq
{\cal A}({\bf p}, {\bf p}^\prime)_{\lambda \,\lambda^\prime}
\sim
\delta^3({\bf p}+{\bf p}^\prime)\,
\xi^+_\lambda({\bf p}) (A_i\sigma^i)
\eta_{\lambda^\prime}(-{\bf p})\times
F(p, p)_{\lambda \,\lambda^\prime}.
\eq
It is easy to see from here that, due to the arbitrary orientation of $A_i$, $\bf p$ and ${\bf p}^\prime$,
the vanishing of the amplitudes for all $\lambda$, $\lambda^\prime$ requires $F(p, p)_{\lambda \,\lambda^\prime}=0$.
Considering the arbitrary signs $\sigma$, $\sigma^\prime$ in Eq. (\ref{defF}) this further implies
\bq
f_\pm(p, p)=0.
\eq
But it is rather evident from Eq. (\ref{fds}) that this cannot be true for all $p$. For example, a
simple way to see that $f_\pm(p, p)$ do not identically vanish is by assuming $p\rightarrow \infty$, in
which limit one can use
\bq
H^{(2)}_{\nu_\pm} (-p\eta)\sim \frac{e^{ip\eta} }{\sqrt{-p\eta}}.
\eq
This makes the integrals (\ref{fds}) trivial, leading to a nonzero result (independent of $p$).

\section{The source of the problem}

We now provide an explanation for the gauge dependent amplitudes (\ref{a2}). Let us consider a general gauge
transformation $A_\mu \rightarrow A^\prime_\mu=A_\mu+ \partial_\mu \Lambda.$ The usual way to check the gauge
invariance of Eq. (\ref{a2}) is to make an integration by parts and use
\bq
\partial_\mu(\sqrt {-g }\, \bar \psi_f \gamma^{\mu} \psi_i)=0.
\label{fercon}
\eq
However, one can conclude that the gauge variations of the amplitudes are rigorously zero only if one
can ignore the surface terms. This is justified for the contributions from spatial infinity
(all physical fields vanish in this region), but it is not allowed for the contributions from $t\rightarrow
\pm \infty$. One way to eliminate the surface terms is to
decouple the interaction at infinite times. In these conditions after the integration by parts the
gauge variations of the amplitudes reappear as an integral which contains the derivative with respect
to time of the decoupling factor, and the gauge invariance translates into the fact that
for an adiabatic decoupling these variations must vanish. We now apply this procedure to the
amplitudes (\ref{a2}).

We denote the decoupling factors by $h_\varepsilon$, with $\varepsilon>0$ the decoupling parameter.
The usual requirements to be imposed on the functions $h_\varepsilon (\eta)$ are
\bq
\lim_{\eta\rightarrow 0} h_\varepsilon(\eta)=
\lim_{\eta\rightarrow -\infty} h_\varepsilon(\eta)=0,\,\,\, \mbox{$\varepsilon>0$ fixed},
\label{h2}
\\
\lim_{\varepsilon \rightarrow 0}\,h_{\varepsilon}(\eta)=1,\,\,\, \eta<0\mbox{  fixed},
\qquad\qquad\quad\quad\,\,
\label{h1}
\eq
with the adiabatic limit corresponding to $\varepsilon \rightarrow 0$. Note that in this limit
$h_{\varepsilon}^\prime(\eta)\rightarrow 0$ for $\eta<0$ fixed.
The decoupled amplitudes (\ref{a2}) are
\bq
{\cal A}_{i\rightarrow f}(\varepsilon)=-ie \int d^4 x \sqrt {-g }\, h_\varepsilon\, \bar \psi_f
\gamma^{\mu} \psi_i A_
\mu.
\label{ampdec}
\eq
The gauge variations of the amplitudes (\ref{ampdec}) are
\bq
\Delta {\cal A}_{i\rightarrow f}(\varepsilon)=-ie
\int d^4 x \sqrt {-g }\, h_\varepsilon\, \bar \psi_f \gamma^{\mu} \psi_i\, (\partial_\mu \Lambda).
\label{varamp}
\eq
Integrating by parts and using Eq. (\ref{fercon}) one finds
\bq
\Delta {\cal A}_{i\rightarrow f}(\varepsilon)=ie \int d^4 x \sqrt {-g }\, h_\varepsilon^\prime
\, \bar \psi_f \gamma^{0} \psi_i \, \Lambda.
\label{varamp1}
\eq
Formula (\ref{varamp1}) can be obviously applied also in the Minkowski space. In this case, the key fact
is that the purely oscillatory behavior of the
modes $\sim e^{\pm i E_{\bf p} t}$ keeps the integral with respect to $t$ bounded for all
$\varepsilon >0$, which in turn ensures that for an adiabatic decoupling $h_\varepsilon^\prime(t)\rightarrow 0$\,
the integral vanishes, and thus in this limit Eq. (\ref{varamp}) vanishes. As we now show,
a different situation arises in dS space.

We are interested in Eq. (\ref{varamp1}) in the limit $\varepsilon \rightarrow 0$. It is clear that it is
sufficient to focus on the integral with respect to $x^0\equiv \eta\in (-\infty, 0)$. Notice that
a nonzero result can only come from the possible divergences from the integration limits
$\eta\rightarrow -\infty$ or $0$ (the vanishing of $h_\varepsilon^\prime(\eta)$ for $\eta$ fixed makes
irrelevant the finite values of $\eta$). In the first limit, the oscillatory behavior
of the Hankel functions
\bq \qquad
H_{\nu_\pm}^{(2)}(-ip\eta)\sim e^{ip\eta}, \quad  \eta\rightarrow -\infty,
\label{oschan}
\eq
makes the situation essentially identical to that in Minkowski space, so that this limit gives no contribution
in the final result. It thus remains to consider the contributions from $\eta\rightarrow 0$. Ignoring the
dependence on ${\bf x}$ in the integrand in Eq. (\ref{varamp1}), it is convenient to collect the $\eta$-dependent factors
in the following way:
\bq
\sqrt {-g }\, \, \bar \psi_f \gamma^0 \psi_i \sim {\cal F}_{fi}(\eta).
\label{edep}
\eq
The integral with respect to $\eta$
is then of the following form:
\bq
\Delta {\cal A}_{i\rightarrow f}(\varepsilon)\sim \int_{-\infty}^0
d\eta\, h_\varepsilon^\prime(\eta)\,{\cal F}_{fi}(\eta) \Lambda(\eta).
\label{relint}
\eq
In explicitly considering the factors ${\cal F}_{fi}(\eta)$ there are essentially two combinations for the
initial and final wave functions, i.e. $\bar \psi_f \psi_i= \bar u v^\prime$ or $\bar \psi_f \psi_i=\bar u u^\prime$.
Using $\sqrt {-g }\, \gamma^0 \sim (H\eta)^{-3} \gamma^{\hat 0}$ together with Eqs. (\ref{umod}) and (\ref{vmod}) one
finds (note that all powers of $H$ are included)
\bq
{\cal F}_{\bar u v^\prime}(\eta)= \sqrt{p p^\prime}\, \eta
\qquad\qquad\qquad\qquad\qquad\qquad\qquad\qquad\qquad\qquad\qquad
\nonumber
\\
\quad\times\left\{
\sigma^\prime
H_{\nu_+}^{(2)}(-p\,\eta)  H_{\nu_-}^{(2)}(-p^\prime\eta)-\sigma
H_{\nu_-}^{(2)}(-p\,\eta)  H_{\nu_+}^{(2)}(-p^\prime\eta)\right\},\qquad\quad\,\,\,
\label{f1}
\eq
\bq
{\cal F}_{\bar u u^\prime}(\eta)= \sqrt{p p^\prime}\, \eta
\qquad\qquad\qquad\qquad\qquad\qquad\qquad\qquad\qquad\qquad\qquad
\nonumber
\\
\quad\times\left\{
e^{\pi k}\,
H_{\nu_+}^{(2)}(-p\,\eta)  H_{\nu_-}^{(1)}(-p^\prime\eta)+\sigma\sigma^\prime
e^{-\pi k}\,
H_{\nu_-}^{(2)}(-p\,\eta)  H_{\nu_+}^{(1)}(-p^\prime\eta)\right\}.
\label{f2}
\eq
All other combinations lead to factors ${\cal F}_{fi}(\eta)$  that can be
obtained from Eqs. (\ref{f1}) and (\ref{f2}) via complex conjugation.

We now establish the limit\, $\varepsilon \rightarrow 0$ in Eq. (\ref{relint}). Firstly, according to
the observation below Eq. (\ref{oschan}), we can replace the inferior integration limit with some
finite value $\eta_*<0$,
\bq
\lim_{\varepsilon \rightarrow 0}
\int_{-\infty}^0
d\eta\, h_\varepsilon^\prime(\eta)\,({\cal F}_{fi} \Lambda)(\eta)
=\lim_{\varepsilon \rightarrow 0}
\int_{\eta_*}^0
d\eta\, h_\varepsilon^\prime(\eta)\,({\cal F}_{fi}\Lambda)(\eta)\,
\equiv\,
\ell.
\label{inteta}
\eq
The important fact is that the factors ${\cal F}_{fi}(\eta)$ have a well-defined limit for
$\eta\rightarrow 0$ (see below). This allows to consider $\eta_*$ arbitrarily close to zero in Eq.
(\ref{inteta}), from which
\bq
\ell =
({\cal F}_{fi}\Lambda)(\eta \rightarrow 0)\times
\lim_{\varepsilon \rightarrow 0}
\int_{\eta_*}^0
d\eta\, h_\varepsilon^\prime(\eta)
=-({\cal F}_{fi} \Lambda)(\eta \rightarrow 0),
\eq
where the last identity follows from Eqs. (\ref{h2}) and (\ref{h1}). Thus, the integral (\ref{relint})
in the limit $\varepsilon \rightarrow 0$ is
\bq
\lim_{\varepsilon\rightarrow 0}
\Delta {\cal A}_{i\rightarrow f}(\varepsilon) \sim -({\cal F}_{fi} \Lambda)(\eta\rightarrow 0).
\label{finlim}
\eq
The limits for ${\cal F}_{fi}(\eta)$ in Eqs. (\ref{f1}) and (\ref{f2}) can be obtained with
\bq
\quad
H_{\nu}^{(2)}(z)\simeq -H_{\nu}^{(1)}(z) \simeq \frac{i}{\pi}\Gamma(\nu)\, \left(\frac{z}{2}\right)^{-\nu},
\quad z\rightarrow 0.
\label{limhan}
\eq
Using Eq. (\ref{limhan}) one finds
\bq
{\cal F}_{\bar u v^\prime}(\eta\rightarrow 0)
= \frac{2}{\pi \cosh (\pi k)}
\left
\{\sigma^\prime (p^\prime/p)^{ik}-\sigma(p^\prime/p)^{-ik}
\right\},
\label{lf1}
\eq
\bq
\qquad\quad
{\cal F}_{\bar u u^\prime}(\eta\rightarrow 0)=\frac{-2}{\pi \cosh (\pi k)}
\left
\{e^{+\pi k} (p^\prime/p)^{ik}+\sigma \sigma^\prime e^{-\pi k}(p^\prime/p)^{-ik}
\right\}.
\label{lf2}
\eq
Returning to the original expression of the gauge variations (\ref{varamp1}), one can now read from
Eqs. (\ref{edep}) and (\ref{finlim}) that in the adiabatic limit
\bq
\lim_{\varepsilon\rightarrow 0} \Delta {\cal A}_{i\rightarrow f}(\varepsilon)
= -ie \int_{\eta\rightarrow 0} d^3 {\bf x} \, \sqrt {-g }\, \bar \psi_f \gamma^{0} \psi_i\, \Lambda.
\label{gauvar}
\eq
The integral (\ref{gauvar}) is clearly not identically zero. This provides an alternative proof for the gauge
dependence of the amplitudes (\ref{a2}). In the flat space limit $k=m/H\rightarrow \infty$ and, as expected,
the gauge variations vanish. (See Eqs. (\ref{lf1}) and (\ref{lf2}); the term in ${\cal F}_{\bar u u^\prime}$ which
contains the factor $e^{+\pi k}$ vanishes in a distributional sense due to the rapidly oscillatory factor
$\sim (p^\prime/p)^{ik}$.)

A remarkable fact about Eq. (\ref{gauvar}) is that it reproduces the surface term from $\eta\rightarrow 0$
in the gauge variations of the amplitudes that follow after performing the integration by parts in the
undecoupled amplitude (\ref{genamp}), i.e. (using a more geometric notation)
\bq
\lim_{\varepsilon\rightarrow 0}
\Delta {\cal A}_{i\rightarrow f}(\varepsilon)
=
-ie \int_
{\eta\rightarrow 0}
d^{\,3} {\Sigma}_\mu \bar \psi_f
\gamma^{\mu} \psi_i \,\Lambda.
\eq
Thus, although we have decoupled the interaction at $t\rightarrow \infty$, the
problematic surface term from this limit reemerges for an adiabatic decoupling.

Let us stress that the key piece in the calculation above is the existence of the limits
${\cal F}_{fi}(\eta\rightarrow 0)$. In the Minkowski space, a similar calculation would
fail due to the oscillatory form of the modes, for which the analogous limits do not exist. One
should also stress that the oscillatory behavior of\, ${\cal F}_{fi}(\eta\simeq 0)$\, in
Eqs. (\ref{f1}) and (\ref{f2}) does not disappear due to the modes themselves, which remain
oscillatory in the infinite future. Using Eq. (\ref{limhan}) one finds that in this limit both
the $u$, $v$ modes contain components that oscillate as
\bq
\qquad \sim \eta^{-\nu_\mp}\sim e^{\mp i m t}, \quad t\rightarrow \infty.
\label{oscfac}
\eq
It is the particular combination of the oscillatory terms in the bilinear form $\bar \psi_f \gamma^{0} \psi_i$
that eliminates the oscillatory behavior of ${\cal F}_{fi}(\eta\simeq 0)$. A direct way to see this is to note
that in Eqs. (\ref{f1}) and (\ref{f2}) the arguments $\nu_\pm=\frac{1}{2}\pm ik$ are always paired as
$H_{\nu_+} H_{\nu_-}$, which implies that in all cases
${\cal F}_{fi}(\eta\rightarrow 0) \sim \eta^{1-\nu_+-\nu_-}=1$.

Another key feature is that the frequencies (\ref{oscfac}) do not depend on the comoving momentum ${\bf p}$, which
is what actually eliminates the oscillatory behavior of ${\cal F}_{fi}(\eta\simeq 0)$\, for $all$ initial and final
momenta ${\bf p}$ and ${\bf p}^\prime$. The fact that the evolution of the modes at late times becomes independent
of the comoving momentum is obviously a consequence of the infinite expansion of space at $t\rightarrow \infty$.
Thus, from a physical point of view, the gauge dependence noted here can be attributed to this fact.

One can ask at this point whether our conclusion depends on the specific exponential form of the scale factor
in dS space $a(t)=e^{Ht}$, or it is sufficient for the scale factor to diverge at $t\rightarrow \infty$. The
answer can be obtained by looking at the Dirac equation in a flat FRW spacetime with an arbitrary scale factor
$a(t)$. In conformal coordinates, the equation reads (see e.g. Ref. \cite{lyth})
\bq
(a^{-1} i\gamma^\mu \partial_\mu -m) \psi =0,
\eq
where a factor $a^{-3/2}$ was extracted from $\psi$. Considering a plane wave solution
$\psi(\eta, {\bf x}) \sim w(\eta)\, e^{\pm i{\bf p}\cdot {\bf x}}$, a divergent scale factor
$a\rightarrow \infty$ implies (notice the derivative with respect to $t$)
\bq
(i\gamma^0 \partial_t -m) w=0,
\label{eflat}
\eq
which is identical to the equation in flat space for a particle with ${\bf p}=0$. The general solution of
Eq. (\ref{eflat}) is a superposition of upper components $\sim e^{-imt}$ and lower components $\sim e^{imt}$,
which are precisely the oscillatory factors in Eq. (\ref{oscfac}). This\, shows that the same problem will\,
appear irrespective of the form in which $a(t)$ diverges at $t\rightarrow \infty$.

Note, however, that we assumed until now that the Dirac field is massive. The massless Dirac field
is conformally invariant, and for such fields the dependence on ${\bf p}$ survives via $w(\eta)\sim e^{\mp i p\eta}$.
If the expansion of space is sufficiently slow so that in the infinite future
$\eta\rightarrow \infty$, the situation is essentially similar to that in Minkowski space, and the gauge
dependence will not appear. On the other hand, if the expansion is fast enough so that $\eta\rightarrow
\eta_{max}<\infty$, the modes themselves ``freeze out'' at late times, and the picture is practically the same as
for the massive fields.

Finally, it is clear from the above observations that the source of the problem is the infinite expansion of
space at $t\rightarrow \infty$. It is then also clear that if the physical process of interest is sufficiently
localized in time (e.g., the case of laboratory experiments in a slowly expanding universe), the infinite
expansion can be ignored, and the problem will not appear.

\section{A few more observations}

One could suspect that the gauge dependence of the amplitudes (\ref{a2}) has something to do with the
``unappropriate'' choice of the $out$ modes. However, this is not the case.
If\, we choose a different set of $out$ modes, the amplitudes will be given by a sum over
$S$-matrix elements between states defined by the $in$ modes (like the amplitudes (\ref{a2})), multiplied
by the Bogolubov coefficients which combine the two sets of modes \cite{birr}. One could then hope that the gauge
variations due to the various terms will cancel among themselves in the sum. Unfortunately, this cannot generally
be so, as one can easily check that the cancelation does not happen for the amplitudes considered here
(see Ref.~\cite{nico}).

Another source of the problem could be the $in$-$out$ form of the amplitudes. Such amplitudes are known to be
problematic for the eternally expanding dS space, for which there are ambiguities in the definition of the
asymptotic $out$ states \cite{bous, boya}. The more appropriate approach for eternally expanding spaces is
to extract the measurable quantities from the expectation values of Heisenberg operators in some initial state
of the field using the $in$-$in$ formalism \cite{wein}. As a side remark, there exists
indeed a significant number of\, QED calculations in the expanding dS space based on this formalism, and many
of them involve the photon propagator (see e.g. Refs.~\cite{kahy1, prok1, prok2,  leon2, kahy}). Typically,
these calculations are done with the propagator in a particular gauge, which naturally raises the question of the
gauge invariance of these results. To our knowledge, this property has not yet been firmly established. It might
be relevant in this context to note that, by causality, the expectation values of Heisenberg operators at finite
times will not involve the problematic limit $t\rightarrow \infty$. This could be seen as an indication that in
the $in$-$in$ formalism the problem of gauge dependence will not appear.

Finally, the problem noticed here could simply be a perturbative artefact. It became clear in recent
years that a variety of infrared divergences in Feynman diagrams in the expanding dS space can
be eliminated by resummation techniques (see e.g. Refs.~\cite{seer, yous1}). It could similarly happen
that a nonperturbative calculation will lead to gauge invariant amplitudes.

\section*{Acknowledgments}

One of us (N.N.) thanks Professor Woodard Richard for discussions on the subject.

\end{document}